\documentclass[
preprint,
showpacs,
showkeys
]{revtex4-1}
\usepackage{amsthm}
\theoremstyle{plain}
\newtheorem*{theorem*}{Theorem}

\newtheorem*{definition*}{Definition}

\newtheorem*{lemma*}{Lemma}

\usepackage{braket}
\usepackage{delarray}
\usepackage{here}

\usepackage{txfonts}

\usepackage[dvipdfmx]{graphicx}
\usepackage{bm}
\usepackage{color}
\usepackage{ulem}

\newcommand{\be}{\begin{eqnarray}}
\newcommand{\ee}{\end{eqnarray}}
\newcommand{\ba}{\begin{array}}
\newcommand{\ea}{\end{array}}
\newcommand{\bmat}{\left(\begin{array}}
\newcommand{\emat}{\end{array}\right)}

\begin{document}
\title{
Small-quench Loschmidt dynamics near quantum critical points
}

\author{Kohei Kobayashi}

\affiliation{
Department of Physics, Institute of Science Tokyo, 2-12-1 Ookayama, Meguro-ku, Tokyo, Japan
}

\begin{abstract} 

We study the short-time Loschmidt dynamics after a small sudden quench near a quantum critical point. We show that the initial quadratic growth of the Loschmidt rate function is governed by the variance of the quench operator per system size. For a local quench operator, this variance density is exactly equal to the spatial sum of the equal-time connected two-point correlation function in the initial ground state. This relation connects the early-time Loschmidt response to static critical correlations.
Assuming power-law correlations at criticality, we classify the finite-size scaling of the short-time coefficient by the scaling dimension of the quench operator. In one dimension, the coefficient is finite, logarithmically enhanced, or algebraically enhanced with system size. We illustrate this operator dependence in the transverse-field Ising chain, where transverse-field and longitudinal-field quenches couple to different critical operators. We also discuss the first correction beyond the quadratic regime using the fourth cumulant of the post-quench Hamiltonian.


\end{abstract}
\date{\today}
\maketitle

\section{Introduction}

Quantum critical points provide a natural setting in which static correlations and real-time dynamics are strongly intertwined. At zero temperature, a continuous change of a control parameter can drive a qualitative change of the ground state, causing a quantum phase transition (QPT) \cite{Sachdev, Hertz, Millis, Vojta, Sondhi, Osterloh}.
Such transitions are driven by quantum fluctuations associated with noncommuting terms in the Hamiltonian, and they play an important role in many areas of quantum science, including quantum computation  \cite{Lloyd}, quantum metrology \cite{Giovannetti},
and quantum control \cite{Hegerfeldt, Cimmarusti}.

Near a quantum critical point, the closing of the energy gap and the growth of correlation length make the system highly sensitive to perturbations. 
This sensitivity is useful in some contexts, but it also implies dynamical fragility.
Even a small change of the Hamiltonian can produce a pronounced many-body response, 
such as defect production \cite{Polkovnikov}, the Kibble-Zurek mechanism \cite{Zurek, Dziarmaga}, and quench dynamics \cite{Rigol, Calabrese}.
Therefore, it is a central problem to understand this response in nonequilibrium quantum many-body physics.

A standard way to study quantum critical dynamics is to perform a sudden quench. 
One initially prepares the system in the ground state $|\psi_0\rangle$
of an initial Hamiltonian $H_i$, 
suddenly changes a control parameter, and then studies the time evolution under 
the post-quench Hamiltonian $H_f$.
An important question in this setting is whether the nonequilibrium dynamics 
can show sharp behavior in time. 
This idea leads to the concept of dynamical quantum phase transitions (DQPTs), which are characterized by nonanalytic behavior at certain critical times \cite{Heyl2013,Heylrev}. 
To study DQPTs, one introduces the Loschmidt amplitude
\begin{equation}
G(t)=\langle\psi_0|e^{-itH_f}|\psi_0\rangle,
\end{equation}
which measures the overlap between the initial state and the time-evolved state. 
The corresponding Loschmidt echo is defined as
\begin{equation}
\mathcal{L}(t)=|G(t)|^2.
\end{equation}
From this quantity, one can define the rate function:
\begin{equation}
f(t):=-\frac{1}{N}\log \mathcal{L}(t),
\end{equation}
where $N$ denotes the total number of lattice sites 
(below, we will use $L$ for the linear system size).
Since the Loschmidt amplitude is formally similar to a partition function, 
$f(t)$ is often regarded as a dynamical analogue of a free-energy density \cite{Heyl2013}. 
The Loschmidt echo, also closely related to fidelity, has been extensively studied as a probe of quantum reversibility, sensitivity to perturbations, and many-body dynamics \cite{Gorin, Wimberger}. 
Its decay has also been related to Lyapunov-type behavior, including in the initial-time regime \cite{Fine}. 
In recent years, DQPTs have been studied from many viewpoints, 
including dynamical order parameters \cite{order1, order2, order3, order4}
and topological aspects \cite{topology}, and
scaling and universality \cite{scaling1, scaling2}. 
Several phenomena related to DQPTs have also been observed experimentally in systems such as trapped ions \cite{Jurcevic,Zhang}, Rydberg atoms \cite{Bernien}, and ultracold atoms \cite{Flaschner, Tian, Smale}.
At the same time, recent advances in experimental time resolution and coherent control make it increasingly possible to resolve real-time dynamics after a quench. 
This motivates a more detailed understanding of which aspects of short-time dynamics are universal and which depend on microscopic details.

 Related studies of near-critical quantum quenches have shown that the post-quench dynamics and operator content can depend sensitively on the operator that drives the quench \cite{Delfino1, Delfino2}. 
These results show that nonequilibrium dynamics near a critical point is not determined by the initial critical Hamiltonian alone, but can also depend on the direction of the quench in operator space.

Despite these developments, one basic question remains unclear:
what determines the initial growth of the Loschmidt rate function near a quantum critical point?
Much of the literature on DQPTs focuses on nonanalytic behavior, while the universal structure of the short-time regime is less understood. 
 General quantum speed limits (QSLs) provide bounds on the time required for quantum states to evolve under specified dynamical constraints \cite{MT, Deffner}. However, in extended many-body systems, such bounds do not in general directly resolve 
how long-distance critical correlations and the choice of quench operator determine the early-time Loschmidt dynamics.

The fidelity approach to QPTs was developed extensively in early works by Zanardi and coworkers,
where ground-state overlaps, fidelity susceptibility, and the enhancement of Loschmidt-echo decay near criticality were investigated \cite{Zanardi, Quan, Gu, Venuti}.
These studies established the strong sensitivity of fidelity-based quantities to quantum criticality. 
The question we address here is complementary: how is the initial Loschmidt response determined by the particular local operator that generates the quench?

In the present study, we analyze small sudden quenches near quantum critical points from an operator-resolved point of view. 
The short-time expansion of the Loschmidt echo is controlled by the energy variance of the post-quench Hamiltonian. 
Our aim is to show how this general fact becomes a concrete 
finite-size scaling statement near criticality once the quench direction in operator space is specified.

Our first result is that
for a quench $H_f=H_i+\delta\lambda V$ from the ground state of $H_i$,
the initial quadratic coefficient of the Loschmidt rate function 
is given by the variance density of the quench operator. 
When $V$ is local, this variance density is exactly the spatial sum of the equal-time connected two-point correlation function in the initial ground state. 
Combining this identity with standard critical scaling, 
we obtain a finite-size classification of the short-time coefficient in terms of the scaling dimension of the operator that drives the quench.

We illustrate this classification in the critical transverse-field Ising chain. 
A transverse-field quench probes the energy-density sector and gives a finite short-time coefficient, 
whereas a longitudinal-field quench probes the order-parameter sector and produces an algebraically enhanced coefficient. 
Thus the early-time Loschmidt response is not determined by the critical point alone, 
but also by the operator direction of the quench.

Finally, we discuss the first correction beyond the quadratic regime. 
Using the cumulant expansion, we show that the fourth cumulant of the post-quench Hamiltonian contains both noncommutative contributions involving the initial Hamiltonian and the intrinsic fourth cumulant of the quench operator. This separation clarifies which part of the quartic correction is governed by static higher-order correlations and which part is generated by the noncommuting structure of the quench.

\section{Short-time Loschmidt dynamics}

\subsection{Short-time expansion and correlation scaling}

We consider a small sudden quench described by
\begin{equation}
H_f = H_i + \delta \lambda V,
\end{equation}
where $V$ is the quench operator 
and $\delta \lambda$ is a small real parameter.

To obtain the short-time behavior of $f(t)$, 
we expand the time-evolution operator:
\begin{equation}
e^{-itH_f}=1-itH_f-\frac{t^2}{2}H_f^2+O(t^3).
\end{equation}
Taking the expectation value in the initial state, we obtain
\begin{equation}
G(t)=1-it\langle H_f\rangle-\frac{t^2}{2}\langle H_f^2\rangle +O(t^3),
\end{equation}
where we define
$\langle A\rangle:=\langle \psi_0|A|\psi_0\rangle$.
For notational simplicity, all expectation values, variance, 
and connected correlations below are taken with respect to the initial ground state $|\psi_0\rangle$.
Multiplying by the complex conjugate gives
\begin{equation}
\mathcal{L}(t)=G^*(t)G(t)=1-\mathrm{Var}(H_f)t^2+O(t^4),
\end{equation}
where 
$\mathrm{Var}(A):=\langle A^2\rangle-\langle A\rangle^2$.
Since \(\mathcal{L}(t)=\mathcal{L}(-t)\), odd powers of 
\(t\) vanish in the expansion of \(\mathcal{L}(t)\).
Using $\log(1-x)=-x+O(x^2)$, we immediately find
\begin{equation}
f(t)=\frac{\mathrm{Var}(H_f)}{N}t^2+O(t^4).
\end{equation}

Since $|\psi_0\rangle$ is an eigenstate of $H_i$,
the variance of $H_f$ in the initial state is given by the quench term. 
Indeed,
\begin{equation}
H_f-\langle H_f\rangle
=(H_i-E_0)+\delta \lambda(V-\langle V\rangle),
\end{equation}
where $E_0:=\langle H_i \rangle $ and $(H_i-E_0)|\psi_0\rangle=0$. 
It follows that
\begin{equation}
\mathrm{Var}(H_f)=\delta\lambda^2\mathrm{Var}(V).
\end{equation}
Therefore, the short-time expansion of $f(t)$ is
\begin{eqnarray}
\label{shorttime}
f(t)&=&\delta \lambda^2 \chi_V  t^2 +O(t^4), 
\end{eqnarray}
where $\chi_V=\mathrm{Var}(V)/N$ is the variance density of the quench operator.
This relation is exact for any finite system and does not rely on criticality.

We now rewrite the variance density in terms of equal-time correlations. 
Assume that the quench operator is a sum of local operators:
\begin{equation}
V=\sum_x v_{x},
\end{equation}
where $x$ labels lattice sites.
Then
\begin{equation}
\mathrm{Var}(V)=\sum_{x,y}
\left(\langle v_{x}v_{y}\rangle 
-\langle v_{x}\rangle\langle v_{y}\rangle \right).
\end{equation}
This leads to the equal-time connected correlation function:
\begin{equation}
C(x,y):=\langle v_xv_y\rangle-\langle v_x\rangle\langle v_y\rangle.
\end{equation}
Therefore,
\begin{equation}
\mathrm{Var}(V)=\sum_{x,y} C(x,y).
\end{equation}

If the initial state is translation invariant, 
then $C(x,y)$ depends only on the relative distance. Writing
$r:=x-y$, we have $C(x,y)=C(r)$.
For each fixed $x$, summing over all $y$ is the same as summing over all $r$. 
Hence $\sum_y C(x-y)=\sum_r C(r)$.
Substituting this into the previous expression, we obtain
$\mathrm{Var}(V)=\sum_x\sum_r C(r)=N\sum_r C(r)$.
Thus
\begin{equation}
\label{vardensity}
\chi_V =\sum_r C(r).
\end{equation}
Eqs. (\ref{shorttime}) and (\ref{vardensity}) 
are exact identities for a translationally invariant initial state.
This identity shows that the short-time coefficient of 
the Loschmidt rate function is exactly the spatial sum of equal-time connected correlations of the local quench operator.

We now introduce the critical scaling assumption. 
At the critical point, the local operator $v_x$ can be associated, in the long-wavelength description, with one or more scaling operators of the critical fixed point \cite{Sachdev, Francesco}. 
We denote by $\Delta$ the scaling dimension of the leading scaling operator contributing to the long-distance connected correlation function of $v_x$.
Equivalently, $\Delta$ determines the asymptotic decay
\begin{equation}
C(r)\sim |r|^{-2\Delta},
\end{equation}
for $r\to\infty$. 
Thus, $\Delta$ is determined both by the universality class of the critical point and by the particular local operator that generates the quench. For example, in the one-dimensional transverse-field Ising universality class, the energy-density operator has $\Delta=1$, whereas the order-parameter operator has $\Delta=1/8$.
Substituting this asymptotic form into (\ref{vardensity}), 
we find that the variance density is controlled by
\begin{equation}
\chi_V\sim \sum_r |r|^{-2\Delta}.
\end{equation}
In a large system, this sum can be estimated by an integral. 
In dimension $d$, one has
\begin{equation}
\chi_V \sim \int_a^L d^dr\, r^{-2\Delta},
\end{equation}
where $L$ is the linear system size and $a$ is a short-distance cutoff of the order of the lattice spacing. 
Using spherical coordinates, this becomes
\begin{equation}
\chi_V \sim \int_a^L dr r^{d-1-2\Delta}.
\end{equation}
Evaluating this integral gives three cases:

\begin{equation}\label{VarScaling}
\chi_V \sim
\left\{
\begin{array}{ll}
{\rm const}, & 2\Delta>d,\\[1mm]
\log L, & 2\Delta=d,\\[1mm]
L^{d-2\Delta}, & 2\Delta<d .
\end{array}
\right.
\end{equation}

Away from the critical point but still in its vicinity, the power-law form is cut off by the
finite correlation length $\xi$.
More generally, one may write
\begin{equation}
C(r)\sim r^{-2\Delta}\Phi(r/\xi),
\end{equation}
where $\Phi(x)$ decays rapidly for $x\gg 1$.
In this case, the infrared cutoff in the above
estimate is not $L$ alone, but $L_{\rm eff}\sim \min(L,\xi)$.
Thus the critical finite-size
forms in Eq. (21) describe the regime in which $L$ is smaller than or comparable to
$\xi$, while away from criticality the same expressions are cut off by replacing $L$ with $\xi$ when $\xi<L$.

Combining (\ref{shorttime}) and (\ref{VarScaling}), 
we obtain the general short-time classification:

\begin{equation}
\label{fScaling}
f(t)\sim \delta \lambda^2 t^2\times 
\left\{
\begin{array}{ll}
{\rm const}, & 2\Delta>d,\\[1mm]
\log L, & 2\Delta=d,\\[1mm]
L^{d-2\Delta}, & 2\Delta<d .
\end{array}
\right.
\end{equation}

It is also useful to express this result in terms of a time scale. 
Let $f_*>0$ be a fixed threshold value, and define
\begin{equation}
\tau_{f_*}:=\inf\{t>0:f(t)\ge f_*\}.
\end{equation}

Combining these,
we find the classification of threshold time scale:

\begin{equation}
\label{tauscaling}
\tau_{f_*}(V)
\sim
\frac{\sqrt{f_*}}{|\delta \lambda|}\times 
\left\{
\begin{array}{ll}
{\rm const}, & 2\Delta>d,\\[1mm]
1/\sqrt{\log L}, & 2\Delta=d,\\[1mm]
L^{-(d-2\Delta)/2}, & 2\Delta<d .
\end{array}
\right.
\end{equation}

The estimate in Eq. (\ref{tauscaling}) should be understood
as a result within the initial quadratic short-time regime.
More precisely, if $\tau_{\rm quad}$ denotes the 
time scale at which higher-order cumulants become comparable to the quadratic term, 
the threshold-time scaling is valid only when
$\tau_{f_*}\ll \tau_{\rm quad}$.
In other words, the threshold $f_*$ must be chosen sufficiently 
small so that the $f(t)$ reaches $f_*$ before quartic and higher-order terms become important.
Therefore, Eq. (\ref{tauscaling}) should not be interpreted as a universal fixed-threshold asymptotic result in all scaling limits.
Rather, Eq. (\ref{tauscaling}) describes how the initial curvature is enhanced by finite-size effects within the short-time regime dominated by the quadratic term.

Equation (\ref{tauscaling}) shows that the approach to 
a quantum critical point can strongly shorten the time scale of the initial Loschmidt decay. 
When $2\Delta>d$, the connected correlation function decays sufficiently fast at long distances. 
As a result, its spatial sum remains finite in the thermodynamic limit, and the threshold time stays of order $\sqrt{f_*}/|\delta \lambda|$.
In this case, the initial decay is not strongly enhanced by long-distance critical correlations. 
At the point $2\Delta=d$, the variance density grows logarithmically, and the threshold time is reduced by a factor $1/\sqrt{\log L}$. 
For a sufficiently relevant quench operator, $2\Delta<d$, the variance density grows algebraically with system size, and the threshold time is correspondingly shortened as a power law in $L$.

Thus, the early-time Loschmidt response near a quantum critical point is operator dependent. 
Criticality alone does not determine the initial decay rate; what matters is the scaling dimension of the operator that drives the quench. 
A quench along an operator with slowly decaying correlations produces a large variance density and therefore a rapid initial decay of the LE. 
In this sense, Eq. (\ref{tauscaling}) gives a simple criterion for dynamical fragility near criticality.

\subsection{Application to the transverse-field Ising model}

We consider the family of transverse-field Ising Hamiltonians
\cite{Pfeuty}:
\begin{equation}
H(g)=
-J\sum_j\sigma_j^x\sigma_{j+1}^x
-Jg\sum_j\sigma_j^z,
\end{equation}
where we set $\lambda=g$.
The three Pauli matrices are $\sigma^k_j$($k=x, y, z$), $j=1,\cdots,N$, and 
$\sigma^k_j$ acts on the $j$th site.
The critical point is located at $g=1$. 
We take the initial Hamiltonian to be
$H_i=H(1)$.
A small transverse-field quench is introduced by changing
$g=1$ to $g=1+\delta g$:
\begin{eqnarray}
    H_f=H(1+\delta g)=H_i-J\delta g\sum_j\sigma_j^z,
\end{eqnarray}
where $|\delta g|\ll 1$.
Thus the quench operator is, up to an overall constant,
\begin{equation}
V_z=\sum_j\sigma_j^z .
\end{equation}

At criticality, $\sigma_j^z$ couples to the energy-density operator of the Ising conformal field theory, whose scaling dimension is $\Delta=1$ \cite{Francesco}.
Hence $2\Delta=2>1$.
From (\ref{VarScaling}) and (\ref{fScaling}), 
the variance density remains finite, $\chi_V \sim O(1)$ and therefore
$f(t)\sim \delta g^2 t^2$.
Thus the threshold time remains of order 
\begin{equation}
\tau_{f_*}(V_z)\sim  \frac{\sqrt{f_*}}{|\delta g|},
\end{equation}
even at the critical point.

Next we consider a longitudinal-field quench,
\begin{equation}
H_f=H_i-J\delta h\sum_j\sigma_j^x,
\end{equation}
where we set $\lambda=h$ and the quench operator is proportional to
\begin{equation}
V_x=\sum_j \sigma_j^x.
\end{equation}
Here $\sigma_j^x$ corresponds to the Ising order-parameter field, 
whose scaling dimension is $\Delta=1/8$ \cite{Francesco}.
Hence $2\Delta=1/4<1$, the threshold time is estimated by
\begin{equation}
\tau_{f_*}(V_x) \sim \frac{\sqrt{f_*}}{|\delta h|L^{3/8}}.
\end{equation}

These two examples show that 
the same critical point can display different early-time Loschmidt dynamics. 
The difference comes from the direction of the quench. 
A transverse-field quench probes the energy-density sector and gives a finite short-time coefficient. 
A longitudinal-field quench probes the order-parameter sector and produces a strong algebraic enhancement with system size.

\section{Beyond the quadratic regime}

\subsection{Higher-order cumulants and the validity of the quadratic regime}

The short-time behavior found above is controlled by the quadratic term.
This term gives the initial curvature, but it is not clear
how long the quadratic approximation remains valid. 
To answer this question, we now study the first correction beyond the quadratic regime.

We expand the logarithm of the Loschmidt amplitude 
$G(t)=\langle \psi_0|e^{-itH_f}|\psi_0\rangle$ in cumulants:
\begin{equation}
\log G(t)=\sum_{n=1}^{\infty}\frac{(-it)^n}{n!}\kappa_n(H_f),
\end{equation}
where $\kappa_n(H_f)$ is the $n$th cumulant of $H_f$. 
Since $f(t)=-\frac{2}{N}\mathrm{Re}\{\log G(t)\}$,
only even cumulants contribute.
Therefore,
\begin{equation}
\label{fk4}
f(t)=c_2t^2-\frac{c_4}{12}t^4+O(t^6),
\end{equation}
where the two coefficients are
$c_2:=\kappa_2(H_f)/N$ and $c_4:=\kappa_4(H_f)/N$.

As shown in Sec. 2.1, the second cumulant is 
\begin{equation}
\kappa_2(H_f)
=\delta \lambda^2 \kappa_2(V)
=\delta \lambda^2 \mathrm{Var}(V).
\end{equation}
Thus the quadratic coefficient is
\begin{equation}
c_2=\delta \lambda^2 \chi_V.
\end{equation}

By contrast, the fourth cumulant is more complex.
To make this point explicit, we introduce the centered operators
$A=H_i-E_0$, and $B=V-\langle V\rangle$.
Then
\begin{equation}
H_f-\langle H_f\rangle=A+\delta \lambda B,
\end{equation}
and $A|\psi_0\rangle=0$. 
In expanding the fourth power, the order of the operators must be kept. Terms in which
$A$ appears at the far right vanish because $A|\psi_0\rangle=0$,
and terms in which
$A$ appears at the far left vanish because $\langle\psi_0|A=0$.
Keeping only the surviving
ordered products gives

\begin{eqnarray}
\label{kappa4}
\kappa_4(H_f)
&=& \langle (H_f-\langle H_f\rangle)^4\rangle
-3\langle(H_f-\langle H_f\rangle)^2 \rangle^2 \nonumber \\
&=&  \langle (A+\delta \lambda B)^4\rangle
-3\langle(A+\delta \lambda B)^2 \rangle^2 \nonumber  \\
&=& \delta \lambda^2\langle B A^2 B\rangle
+\delta \lambda^3\left(\langle B A B^2\rangle+\langle B^2 A B\rangle\right)\nonumber \\
&&+\delta \lambda^4\left(\langle B^4\rangle
-3\langle B^2\rangle^2
\right).
\end{eqnarray}

This expression separates the quartic coefficient into three different contributions; a noncommutative term generated by the action of $H_i$,
a mixed noncommutative
term, and the intrinsic fourth cumulant of the quench operator.

For the first term, we find 
\begin{eqnarray}
A B|\psi_0\rangle
&=&  (H_i-E_0)(V-\langle V\rangle)|\psi_0\rangle  \nonumber \\
&=&
[H_i,V]|\psi_0\rangle.
\end{eqnarray}
Therefore, 
\begin{equation}
\langle B A^2 B\rangle=\|AB|\psi_0\rangle\|^2=\|[H_i, V]|\psi_0\rangle\|^2,
\end{equation}
where $\||x\rangle\|^2=\langle x|x\rangle$ is the Euclidean norm.
It quantifies the noncommutativity 
between $H_i$ and $V$ in the initial state.

We consider the second term in Eq. (\ref{kappa4}):
\begin{equation}
\langle BAB^2\rangle+\langle B^2AB\rangle
=2{\rm Re}\langle BAB^2\rangle.
\end{equation}
In the eigenbasis of $H_i$, it is written as
\begin{equation}
2{\rm Re}
\sum_n
(E_n-E_0)
\langle \psi_0|B|n\rangle
\langle n|B^2|\psi_0\rangle.
\end{equation}

This term is a mixed contribution involving the excitation components generated by $B$ and $B^2$,
weighted by the energy gaps $E_n-E_0$.
Its physical interpretation is less transparent, 
and here we only use it to emphasize that the full quartic coefficient generally contains noncommutative dynamical contributions.

The last term in Eq. (\ref{kappa4})
is the fourth cumulant of the quench operator. 
It measures the non-Gaussian fluctuation of $V$ in the initial ground state. 
When the quench operator is a sum of local operators,
\begin{equation}
V=\sum_x v_x,
\end{equation}
we define $b_x=v_x-\langle v_x\rangle$. 
Then $B:=V-\langle V\rangle=\sum_x b_x$, and the fourth cumulant of the quench operator is
\begin{equation}
\kappa_4(V)
=\langle B^4\rangle-3\langle B^2\rangle^2 .
\end{equation}
Expanding this expression in terms of the local fluctuations gives
\begin{equation}
\kappa_4(V)
=\sum_{x_1,x_2,x_3,x_4}
\langle
b_{x_1}b_{x_2}b_{x_3}b_{x_4}
\rangle_c ,
\end{equation}
with
\begin{eqnarray}
\langle
b_{x_1}b_{x_2}b_{x_3}b_{x_4}
\rangle_c
&=&
\langle
b_{x_1}b_{x_2}b_{x_3}b_{x_4}
\rangle
-
\langle b_{x_1}b_{x_2}\rangle
\langle b_{x_3}b_{x_4}\rangle
\nonumber\\
&-&
\langle b_{x_1}b_{x_3}\rangle
\langle b_{x_2}b_{x_4}\rangle
-
\langle b_{x_1}b_{x_4}\rangle
\langle b_{x_2}b_{x_3}\rangle.
\end{eqnarray}

If the initial state is translationally invariant, this becomes
\begin{equation}
\label{kappa4_density}
\frac{\kappa_4(V)}{N}
=
\sum_{r_2,r_3,r_4}
\langle
b_0 b_{r_2}b_{r_3}b_{r_4}
\rangle_c .
\end{equation}

Therefore, the order-$\delta\lambda^4$
contribution to $\kappa_4(H_f)$, and hence to the
quartic coefficient of $f(t)$, probes the connected four-point correlation function of the local quench operator.

Using Eq. (\ref{fk4}), the quadratic approximation is valid as long as
\begin{equation}
c_2t^2
\gg
\frac{|c_4|}{12}t^4 .
\end{equation}
This gives the time scale where the quadratic approximation is dominant:
\begin{equation}
\label{tau_quad}
\tau_{\rm quad}
:=\sqrt{\frac{12c_2}{|c_4|}}.
\end{equation}
For $t\ll \tau_{\rm quad}$, $f(t)$ is well described by the quadratic form, while
for $t\geq \tau_{\rm quad}$, it is no longer determined solely by the two-point correlation function.

Finally if $H_i$ and $V$ commute, the noncommutative terms in 
Eq. (\ref{kappa4}) vanish and 
\begin{equation}
\kappa_4(H_f)=\delta \lambda^4\kappa_4(V).
\end{equation}
In this special case, the quartic correction is directly 
determined by the integrated connected four-point 
function of the quench operator.

\subsection{Fourth-cumulant contribution for the longitudinal-field quench}

We now apply the cumulant analysis to the critical transverse-field Ising chain. 
We focus on the longitudinal-field quench, because it probes the order-parameter sector and gives the clearest infrared enhancement of the fourth-cumulant contribution. 
The initial and final Hamiltonians are
\begin{eqnarray}
H_i &=& -J\sum_j \sigma_j^x\sigma_{j+1}^x - J\sum_j\sigma_j^z, \\
H_f &=& H_i-J\delta h\sum_j\sigma_j^x,
\end{eqnarray}
and the quench operator is
$V_x=\sum_j\sigma_j^x$, whose scaling dimension is $\Delta=1/8$
\cite{Francesco}.
Since \(2\Delta=1/4<1\), the variance density is algebraically enhanced:
\begin{equation}
\chi_V\sim L^{1-2\Delta}=L^{3/4}.
\end{equation}
Thus, $c_2$ scales as
\begin{equation}
\label{c2scaling}
c_2\sim \delta h^2 L^{3/4}.
\end{equation}

We next derive the scaling of the quartic coefficient 
\begin{equation} 
c_4=\frac{\kappa_4(H_f)}{N}. 
\end{equation} 
For a noncommuting quench, the fourth cumulant of 
$H_f$ contains three contributions,
\begin{equation}
c_4=c_4^{(2)}+c_4^{(3)}+c_4^{(4)}, 
\end{equation} 
where 
\begin{eqnarray} 
c_4^{(2)} &=& \frac{\delta h^2}{N} \langle BA^2B\rangle,  \\
c_4^{(3)} &=& \frac{\delta h^3}{N} \left( \langle BAB^2\rangle + \langle B^2AB\rangle \right),
 \\
c_4^{(4)} &=& \frac{\delta h^4}{N}\kappa_4(V_x). 
\end{eqnarray} 

 We first consider $c^{(2)}_4$. We have
\begin{equation}
\langle BA^2B\rangle = \|[H_i,V_x]|\psi_0\rangle\|^2 .
\end{equation}
Thus this term is controlled by the operator generated by the commutator
$[H_i,V_x]$. Since $V_x=\sum_j\sigma^x_j$, the commutator is also a sum of
local operators. In the scaling description, the commutator with the Hamiltonian
corresponds to a time derivative:
\begin{equation}
[H_i,V_x]\sim i\frac{dV_x}{dt}.
\end{equation}
Therefore, if the local order-parameter field has scaling dimension $\Delta$,
the local operator appearing in $[H_i,V_x]$ has scaling dimension $\Delta+z$, where $z$ is the dynamical exponent. 
For the critical Ising chain, $\Delta=1/8$ and
$z=1$, so this dimension is
$\Delta+z=9/8$.
The quantity $N^{-1}\langle BA^2B\rangle$ can then be estimated as the variance density of this commutator-generated operator. 
Its long-distance part is obtained by
integrating a two-point function of an operator with scaling dimension $\Delta+z$:
\begin{eqnarray}
\frac{1}{N}\langle BA^2B\rangle
&\sim&
\int^L dr r^{-2(\Delta+z)}  \nonumber \\
&\sim& L^{1-2(\Delta+z)}\sim L^{-5/4}.
\end{eqnarray}

The negative exponent means that the infrared contribution is convergent. 
Hence this contribution is not controlled by long distances but by short-distance physics,
and the scaling is
\begin{equation}
\frac{1}{N}\langle BA^2B\rangle \sim O(1).
\end{equation}
Therefore,
\begin{equation}
c_4^{(2)}\sim \delta h^2 O(1).
\end{equation}

Next, we consider $c^{(3)}_4$.
For the critical Ising chain, this term vanishes by the $Z_2$ symmetry of the
initial ground state. The longitudinal quench operator $V_x$ is odd under the
$Z_2$ spin-flip symmetry, whereas $A=H_i-E_0$ is even.
Moreover, in the $Z_2$-symmetric ground state, $\langle V_x\rangle=0$,
and hence $B=V_x-\langle V_x\rangle=V_x$ is also $Z_2$-odd.
Therefore, the operators
$BAB^2$ and $B^2AB$ contain an odd number of $Z_2$-odd factors and are
themselves $Z_2$-odd. 
Their expectation values vanish in the $Z_2$-symmetric
ground state, and consequently
\begin{equation}
c_4^{(3)}=0.
\end{equation}

Finally, we consider $c^{(4)}_4$.
This term is the intrinsic fourth cumulant of the quench operator:
\begin{equation}
c_4^{(4)}=\frac{\delta h^4}{N}\kappa_4(V_x).
\end{equation}
It is directly related to the integrated connected four-point function of the local
order-parameter operator. To see the scaling, 
write $V_x=\sum_j b_j$, where $b_j$
denotes the local fluctuation of $\sigma^x_j$.
Then
\begin{equation}
\frac{\kappa_4(V_x)}{N}=
\sum_{r_2,r_3,r_4}
\langle b_0 b_{r_2} b_{r_3} b_{r_4}\rangle_c .
\end{equation}

For a local scaling operator $b_x$ with scaling dimension $\Delta$, the connected
$n$-point function transforms under a uniform rescaling of all relative coordinates,
$r_i\to b r_i$, as
\begin{equation}
\langle b_0 b_{r_2}\cdots b_{r_n}\rangle_c
\to
b^{-n\Delta}
\langle b_0 b_{r_2}\cdots b_{r_n}\rangle_c .
\end{equation}
Therefore, when the corresponding infrared integral is divergent, the integrated
$n$-th cumulant density scales as
\begin{equation}
\frac{\kappa_n(V)}{N}
\sim
\int^L d^d r_2\cdots d^d r_n\,
\langle b_0 b_{r_2}\cdots b_{r_n}\rangle_c
\sim L^{(n-1)d-n\Delta}.
\end{equation}

For $n=4$, $d=1$, and $\Delta=1/8$,  we obtain
$\kappa_4(V_x)/N\sim L^{5/2}$.
Therefore,
\begin{equation}
c_4^{(4)}\sim \delta h^4 L^{5/2}.
\end{equation}
This is an infrared power-counting estimate. It assumes that the connected
four-point contribution of the order-parameter field is not removed by an additional
symmetry constraint or by an accidental cancellation.

Combining these estimates, the full quartic coefficient scales as
\begin{equation}
\label{c4scaling}
c_4 \sim \delta h^2 O(1) + \delta h^4 L^{5/2}.
\end{equation} 
Using the scaling (\ref{c2scaling}) and (\ref{c4scaling}),
the quadratic time scale is estimated as
\begin{equation} 
\tau_{\rm quad} = \sqrt{\frac{12c_2}{|c_4|}} 
\sim \sqrt{ \frac{\delta h^2 L^{3/4}} {\delta h^2 O(1)+\delta h^4 L^{5/2}} }. 
\end{equation}

Here the scaling should be understood as a finite-size estimate of the cumulant coefficients.
Since the longitudinal field is a relevant perturbation, the limits $\delta h\to 0$ and
$L\to\infty$ need not commute. 
In the regime where
the fourth-cumulant contribution dominates,
\begin{equation}
\delta h^2L^{5/2}\gg 1,
\end{equation}
we obtain
\begin{equation}
\tau_{\rm quad}\sim \frac{1}{|\delta h|L^{7/8}}.
\end{equation}
On the other hand, when the noncommutative contribution
$c_4^{(2)}$ dominates, 
\begin{equation}
\tau_{\rm quad}\sim L^{3/8}.
\end{equation}
For fixed finite-size scaling parameters, these estimates describe a crossover between a
short-distance noncommutative regime and a critical fourth-cumulant regime.
Therefore, $\tau_{\rm quad}$ should be interpreted as the time at which 
the quadratic approximation breaks down, not as the decay time itself.

\section{Conclusion}

We investigated the short-time behavior of the Loschmidt rate function after a small sudden quench near a quantum critical point. 
We showed that the initial quadratic growth is controlled by the variance density of the quench operator. 
For a local quench operator, this variance density is exactly given by the spatial sum of the equal-time connected two-point correlation function in the initial ground state. 
This result establishes a direct connection between early-time Loschmidt dynamics and static critical correlations.

At a critical point, the finite-size scaling of the initial short-time coefficient is determined by the scaling dimension of the local operator that drives the quench. 
Depending on this scaling dimension, the coefficient remains finite, grows logarithmically, or grows algebraically with system size. 
Equivalently, the threshold time for the Loschmidt rate function to reach a fixed value remains finite, is logarithmically reduced, or is algebraically shortened. 
Thus the early-time Loschmidt response near criticality is not determined by criticality alone, but also by the operator direction of the quench.

The transverse-field Ising chain provides a clear example of this operator dependence. 
A transverse-field quench probes the energy-density sector and leads to a finite short-time coefficient. 
By contrast, a longitudinal-field quench probes the order-parameter sector and produces an algebraic enhancement of the initial Loschmidt response with system size. 
This contrast shows that different perturbations at the same critical point can cause different short-time dynamics.

We also discussed the first correction beyond the quadratic regime using the cumulant expansion. 
The decomposition of $\kappa_4(H_f)$ shows that, for a noncommuting quench, the quartic coefficient contains both noncommutative contributions involving the initial Hamiltonian and the intrinsic fourth cumulant of the quench operator. 
In the longitudinal-field quench of the Ising chain, this fourth-cumulant contribution is infrared enhanced, while the full quartic coefficient also contains a noncommutative contribution. 
As a result, the quadratic time scale exhibits a crossover between these two contributions.

It would be interesting to extend this analysis by evaluating the full fourth cumulant explicitly in concrete critical models and by clarifying how the operator-dependent short-time scaling discussed here is connected to later-time nonanalytic behavior associated with DQPTs. These results show that the early-time
Loschmidt response near criticality is not a property of the critical point alone, but of the pair consisting of the critical state and the operator direction along which the system is quenched.

This work was supported by JSPS KAKENHI Grant No. 25H01364.


\end{document}